\documentclass[fleqn,usenatbib]{mnras}

\usepackage{newtxtext,newtxmath}
\usepackage[T1]{fontenc}
\usepackage{makecell}
\usepackage{multirow}
\DeclareRobustCommand{\VAN}[3]{#2}
\let\VANthebibliography\thebibliography
\def\thebibliography{\DeclareRobustCommand{\VAN}[3]{##3}\VANthebibliography}

\usepackage{graphicx}	% Including figure files
\usepackage{amsmath}	% Advanced maths commands
\title[Magneto-Active Environments in Pulsar Binaries]{Magneto-Active Environments in Pulsar Binaries with the MeerKAT Telescope: I. Pulsar sample and their basic properties}

\author[Singha et al.]{Jaikhomba Singha$^{1}$\thanks{E-mail: mjaikhomba@gmail.com},
Dongzi Li$^{2}$\thanks{E-mail: dzli@tsinghua.edu.cn},
Marisa Geyer$^{1}$\thanks{E-mail: marisa.geyer@uct.ac.za},
Maciej Serylak$^{3}$,
Federico Abbate$^{4}$,
Senate Lekomola$^{5}$,
\newauthor
Robert Main$^{6}$,
Andrea Possenti$^{4}$,
Amanda Weltman$^{1}$
\\
$^{1}$High Energy Physics, Cosmology \& Astrophysics Theory Group (HEPCAT), Department of Applied Mathematics, University of Cape Town, South Africa\\
$^{2}$Department of Astronomy, Tsinghua University, Beijing 100084, China\\
$^{3}$SKA Observatory, Jodrell Bank, Lower Withington, Macclesfield, Cheshire, SK11 9FT, UK\\
$^{4}$INAF-Osservatorio Astronomico di Cagliari, via della Scienza 5, 09044 Selargius (Italy)\\
$^{5}$Department of Computer Science, University of Botswana, Gaborone, Botswana\\
$^{6}$McGill University Department of Physics, Trottier Space Institute
}

\date{Accepted XXX. Received YYY; in original form ZZZ}

\pubyear{\the\year{}}

\begin{document}
\label{firstpage}
\pagerange{\pageref{firstpage}--\pageref{lastpage}}
\maketitle

% Abstract of the paper
\begin{abstract}
Eclipsing pulsar binaries and binaries with a high mass companion are ideal systems for studying and understanding the properties of plasma in magneto-ionic environments. In this work, the first paper of a series, we present MeerKAT observations of three pulsar binaries: the high-mass binary PSR J1740$-$3052, the black widow PSR J2051$-$0827 and the redback PSR J1748$-$2446A (Terzan~5A). With the help of MeerKAT's high-sensitivity polarimetric observations, we characterise the properties of these sources, including the linear/circular polarization, dispersion measure (DM), rotation measure (RM) and scattering time. The two eclipsing millisecond pulsars exhibit strong orbital-phase-dependent propagation effects and we observe $\sim$2 eclipses in these systems during our observations. PSR J1740$-$3052 is a binary system with a 231\,d orbital period. The relatively large separation results in a smooth RM variation, enabling us to resolve its variation timescale and constrain the small-scale magnetic structure. A gradual RM variation is observed over $\sim$1500\,s, occurring near periastron. This behaviour implies a magnetic spatial scale of $\sim$0.003\,AU in the companion wind, assuming a relative velocity of $\sim$250\,km\,s$^{-1}$. The linear polarisation intensity profiles of PSR J2051$-$0827 show shape variations as a function of frequency, with a stronger leading component emerging at lower frequencies. We observe signatures of the propagation effect in the polarisation properties of PSR J1748$-$2446A during eclipse ingress and egress. This could arise from Faraday Conversion or multipath propagation of the pulsar signal and requires detailed analysis.

\end{abstract}

% Select between one and six entries from the list of approved keywords.
% Don't make up new ones.

%%%%%%%%%%%%%%%%%%%%%%%%%%%%%%%%%%%%%%%%%%%%%%%%%%

%%%%%%%%%%%%%%%%% BODY OF PAPER %%%%%%%%%%%%%%%%%%

\section{Introduction}
Radio pulsars in binary systems provide unique laboratories for studying magneto-active plasma under extreme physical conditions. The broadband, and highly polarised emission from pulsars acts as a precise probe of the intervening medium, allowing detailed investigations of magnetised environments through measurements of dispersion, scattering and Faraday rotation. In pulsar binaries, these effects can also arise from the magnetic environment of these systems. Such binaries could be (i) eclipsing millisecond pulsars \citep{Roberts2013} with very low-mass companions ($\ll$ 0.1 M$_\odot$), known as black widows or companion masses of a few tenths of solar masses (0.1–0.4 M$_\odot$), known as redbacks or (ii) high mass binary systems. The orbital-phase-dependence of propagation through the companion wind together with the short timescales involved, makes it possible to precisely distinguish between environmental effects to those intrinsic to the pulsar. 

The improvements in telescope sensitivity and bandwidth have advanced the study of rare, highly time-variable and chromatic propagation effects. Extreme plasma lensing of pulsar emission has also been observed in eclipsing binaries such as PSR B1957+20 \citep{Main_etal2018} and PSR J1748$-$2446A \citep{Bilous_etal2019}. These lensing events are sensitive probes of pulsar emission, resolving physical scales of $\sim$\,10 km at the pulsar, and can be used to measure magnetization of intra-binary material and to constrain mass-loss rates. Extreme magneto-environments have been observed in the high-mass binary system PSR B1259$-$63 \citep{Johnston_etal2005} and in PSR J1748$-$2446A, a pulsar with a nearby low-mass companion \citep{Li_etal2023}. The understanding of the magnetic structure, in turn constrains the evolutionary history of the binary system and the eclipse properties of the pulsar emission. 

\begin{table*}
\centering
\caption{Summary of basic properties of the pulsar sample in the present work. The pulsar name based on J2000 coordinates along with the rotational period and orbital period of the system are listed in the first three columns. The observation dates, bands, number of frequency channels and the observation mode are listed in columns 4th to 7th respectively.}
\label{tab:pulsar_properties}
\begin{tabular}{lcccccc}
\hline
Pulsar Name &
$P_{0}$ (s) &
\textit{P$_B$} (days) &
Observation date &
Observation Band &
$N_{\rm chan}$ &
Observation Mode \\
\hline
\multirow{6}{*}{PSR J1740$-$3052}  & \multirow{6}{*}{0.5703} & \multirow{6}{*}{231.03}  & 10/2/2024 &  \multirow{4}{*}{UHF-band}   & \multirow{6}{*}{1024, 4096} & \multirow{6}{*}{Fold, Search} \\
 &  &  &  19/2/2024 &     &  &  \\
 &  &   & 21/2/2024&    &  &  \\
 &  &   & 24/2/2024 &     &  &  \\
    &  &  & 29/2/2024 & \multirow{2}{*}{S1-band}    &  &  \\
    &  &  & 25/9/2024 &    &  &  \\
% Add additional pulsars here
\hline
\multirow{3}{*}{PSR J2051$-$0827}  & \multirow{3}{*}{0.0045} & \multirow{3}{*}{0.1}  & 6/6/2024 & L-band & \multirow{3}{*}{1024} & \multirow{3}{*}{Fold, Search} \\
 & &   & 10/4/2024 & \multirow{2}{*}{UHF-band}  &  &  \\
& &   &  18/6/2024 &  &  &  \\
\hline
\multirow{2}{*}{PSR J1748$-$2446A} & \multirow{2}{*}{0.01156} & \multirow{2}{*}{0.077}  & 27/5/2019 & \multirow{2}{*}{L-band} & \multirow{2}{*}{1024} & \multirow{2}{*}{Search} \\
& &  & 27/2/2020 &  &  &  \\
\hline
\end{tabular}
\end{table*}

A sample of three binary pulsars -- PSR J1740$-$3052 (a high mass binary system), PSR J2051$-$0827 (a black widow) and PSR J1748$-$2444A (a redback) - have been selected for the present study. The selected sources are all expected to be in a magneto-ionic environment. It has been proposed that the magnetic strength of the companion will influence the evolution of the binary system - when coupled to the ablated wind, the braking from the magnetic field could effectively remove orbital angular momentum and thus maintain stable Roche-lobe overflow \citep{Chen_etal2013, Ginzburg_etal2021}. Measurement of magneto-ionic environment in a increasing number of binary systems at different evolutionary stages is necessary to reveal this effect. We also aim to search for, and study plasma lensing events, constraining the size of the pulsar's coherent emission region. In binaries, the proximity of the plasma lens to the companion gives an unprecedented resolution of the pulsar emission regions. However, these techniques have been put into practice only once in studying lensed emission from PSR B1957+20 \citep{Main_etal2018}. Plasma lensing is necessarily chromatic, and the time and frequency scales of lensing, combined with the magnification, determine the lens resolution and the transverse velocity of the companion star's outflow wind. 

The MeerKAT radio telescope \citep{Jonas_etal2016}, thanks to its high high sensitivity, multiple broad observing bandwidths, and high time resolution modes, has opened a new window for pulsar observations in the Southern Hemisphere \citep{Bailes_etal2020}. Its capabilities are particularly well suited to detecting subtle, time-variable magneto-ionic signatures in pulsar binaries across orbital phase. Motivated by these capabilities, we have initiated a MeerKAT binary pulsar observation Open Time proposal aimed at characterizing the magneto-active environments in a sample of pulsar binary systems. The overarching goal of this effort is to quantify the lensing and eclipsing phenomena along with understanding the properties of the magento-ionic plasma within binary systems. In the present paper, we focus on introducing these sources and their basic properties, observed using the MeerKAT telescope. 
\begin{figure}
\centering
\includegraphics[width=0.95\linewidth]{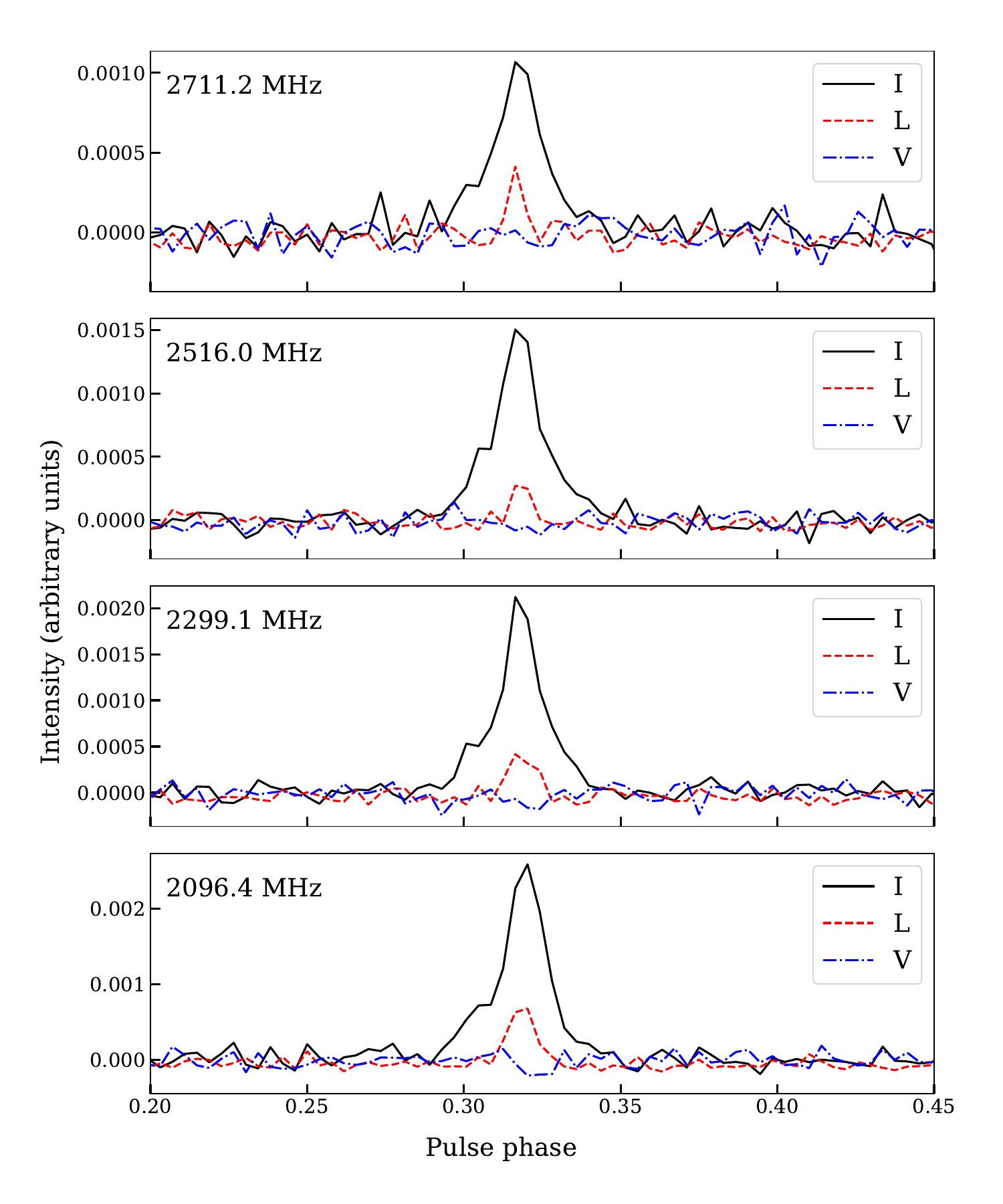}
\caption{The pulse profile of PSR J1740$-$3052 at different frequency channels for 320\,s of sub-integrated time, observed 6 days away from the epoch of periastron. The solid black lines represent the total intensity, the red dashed lines represent linear polarisation and the blue dash-dotted lines represent circular polarisation. There is no visible effect of scatter-broadening at the lower frequency channels of S1-band observations. The pulsar shows very low circular polarisation intensities at all frequency channels.}
\label{fig:profileJ1740}
\end{figure}
\begin{figure*}
\includegraphics[width=0.75\linewidth]{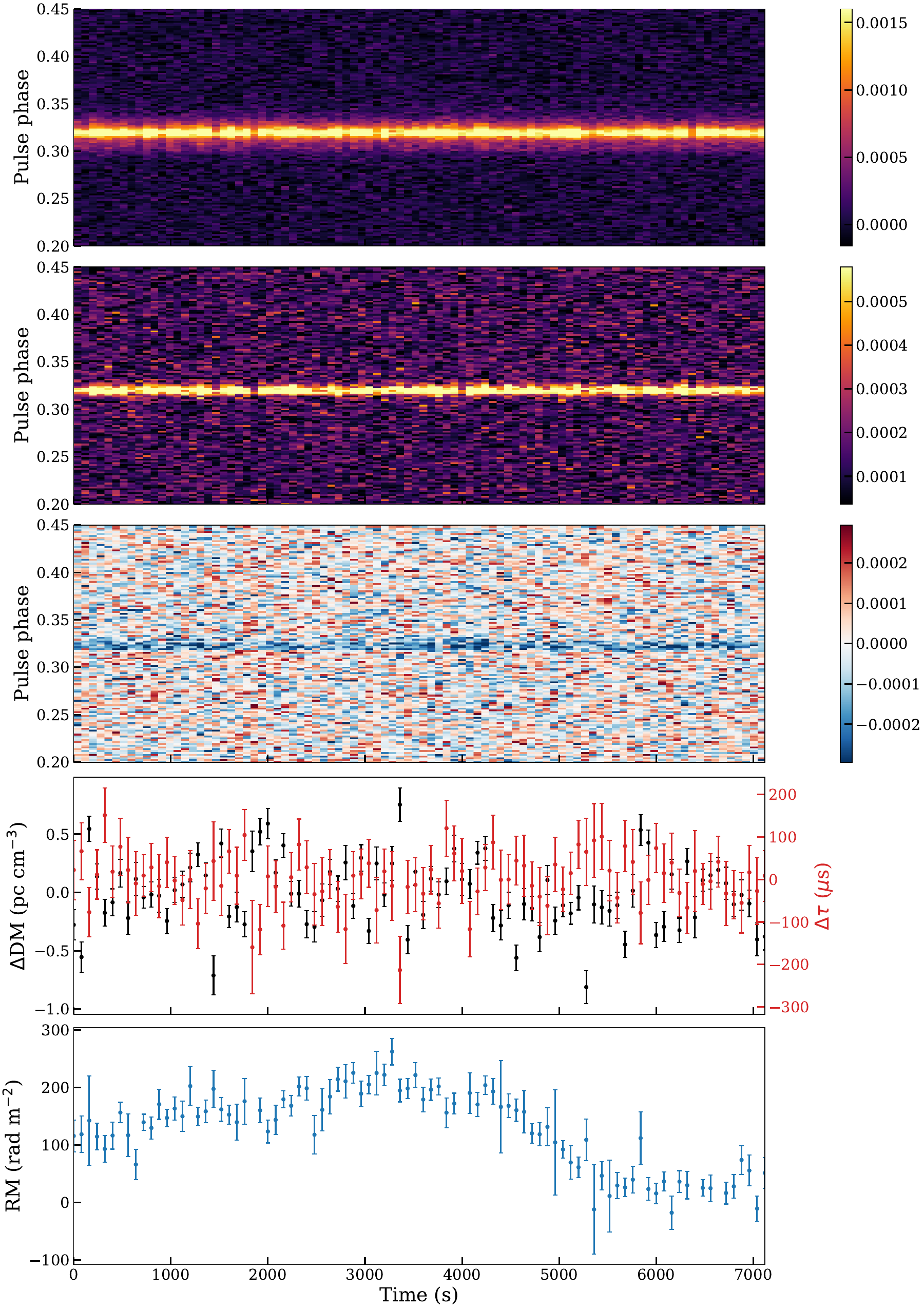} 
\caption{Evolution of total intensity (1st panel/top), linear polarisation intensity (2nd panel), circular polarisation intensity (3rd panel), DM and scattering variation (4th panel) and RM variation (5th panel) as a function of observation time for PSR J1740$-$3052, observed at S1-band with 875 MHz bandwidth the centre frequency of 2405 MHz. There is no visible DM or scattering variation in this pulsar. However, the RM shows a slow variation.}
\label{fig:phasetimeJ1740}
\end{figure*}
The remaining sections of this paper are organised in the following manner. In Section~\ref{sec:observations}, we describe the observations and the techniques used to reduce the data for characterizing and studying effects including dispersion, scattering and Faraday rotation. The results and discussions describing the basic properties of the pulsars in our sample are presented in Section~\ref{sec:results}. Finally, in Section~\ref{sec:summary}, we present the summary of our study and the future work outlining the next steps in this series.

\begin{figure*}
\centering
\includegraphics[width=0.425\linewidth]{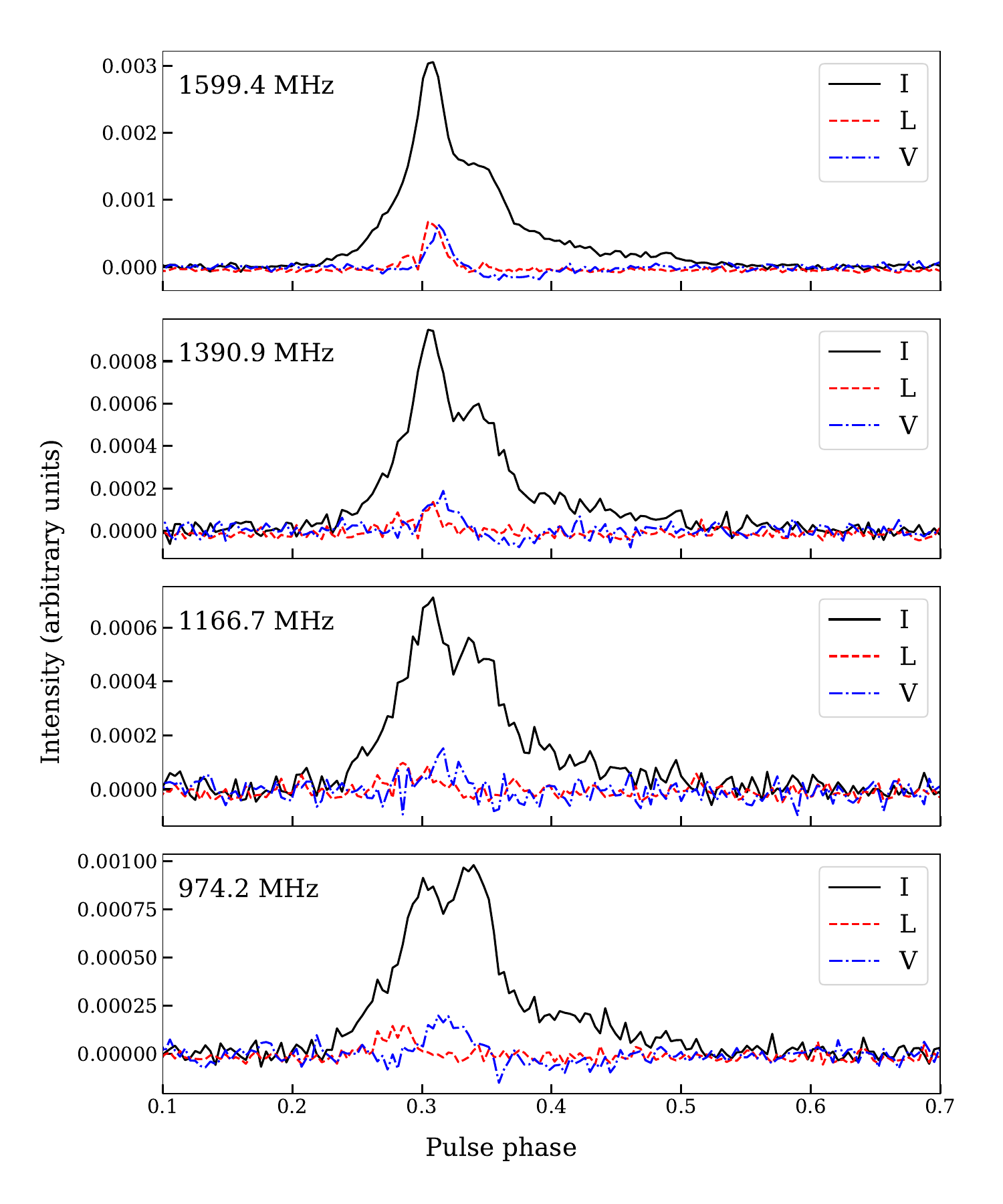}
\raisebox{11mm}{
\includegraphics[width=0.535\linewidth]{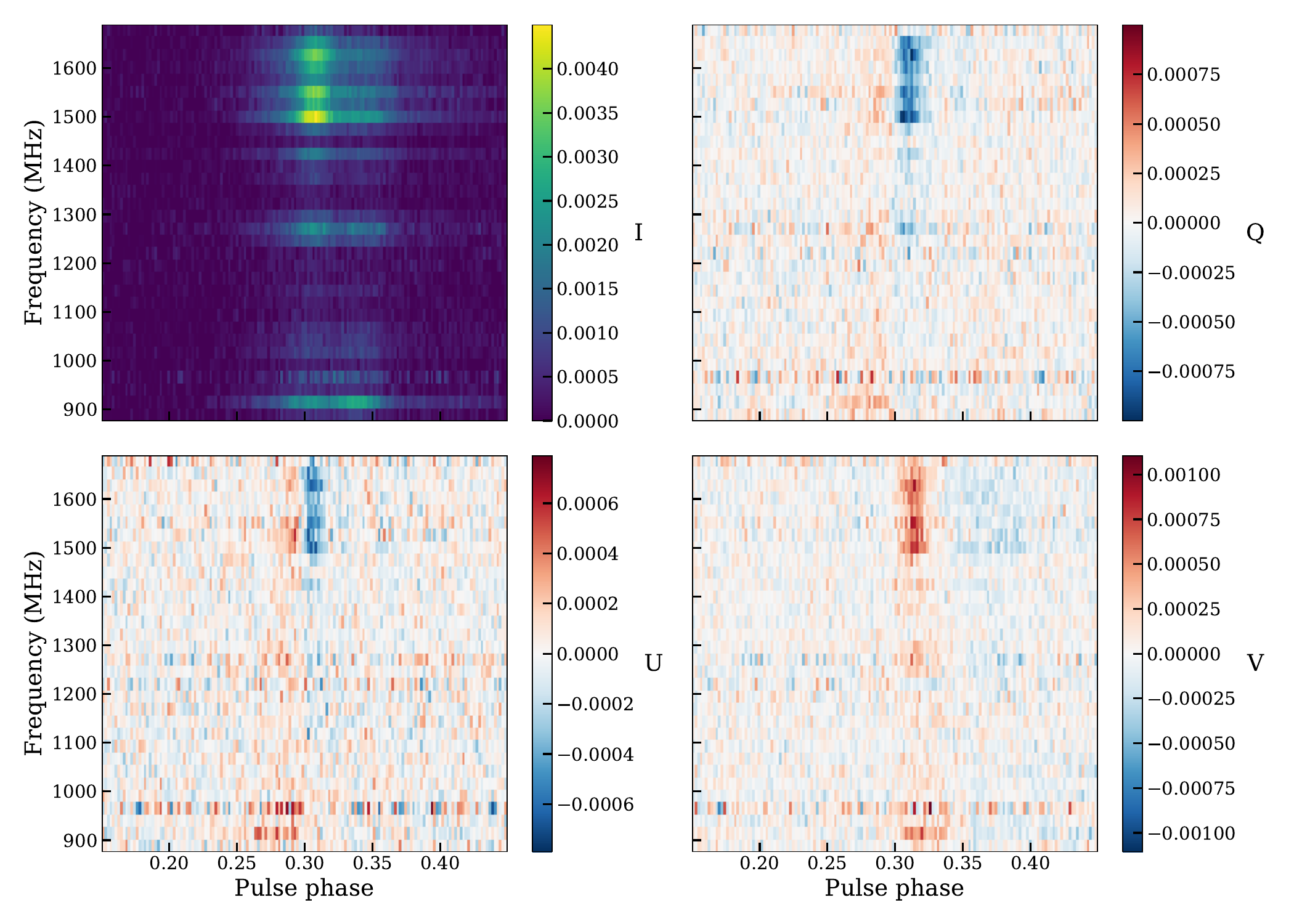}
}

\caption{\textit{Left:} The pulse profile of PSR J2051$-$0827 at different frequency channels for 320\,s of sub-integrated time, far away from the eclipse region. The solid black lines represent the total intensity, the red dashed lines represent linear polarisation and the blue dash-dotted lines represent circular polarisation. The lower frequency channels have lower linear and circular polarisation intensities. We also see clear scintillation in this pulsar as a function of frequency. \textit{Right:} Stokes I, Q, U and V as a function of frequency after being corrected for the value of RM. It is seen that the behaviour of Stokes Q and U changes as a function of frequency, with a stronger leading component emerging at lower frequencies, resulting in overall changes in the linear polarisation intensity profiles as a function of frequency.}
\label{fig:profileJ2051}
\end{figure*}

\section{Observations and Data Reduction}
\label{sec:observations}
Observations of the selected pulsar sample were conducted with the MeerKAT radio telescope, located in the Karoo region of South Africa. The interferometer consists of 64 antennas, each approximately 13.5 m in diameter, and is capable of producing both high-resolution visibilities as well as high time resolution beamformer data products. \citep{Bailes_etal2020, Stappers_etal2018}.  
\begin{figure*}
\includegraphics[width=0.75\linewidth]{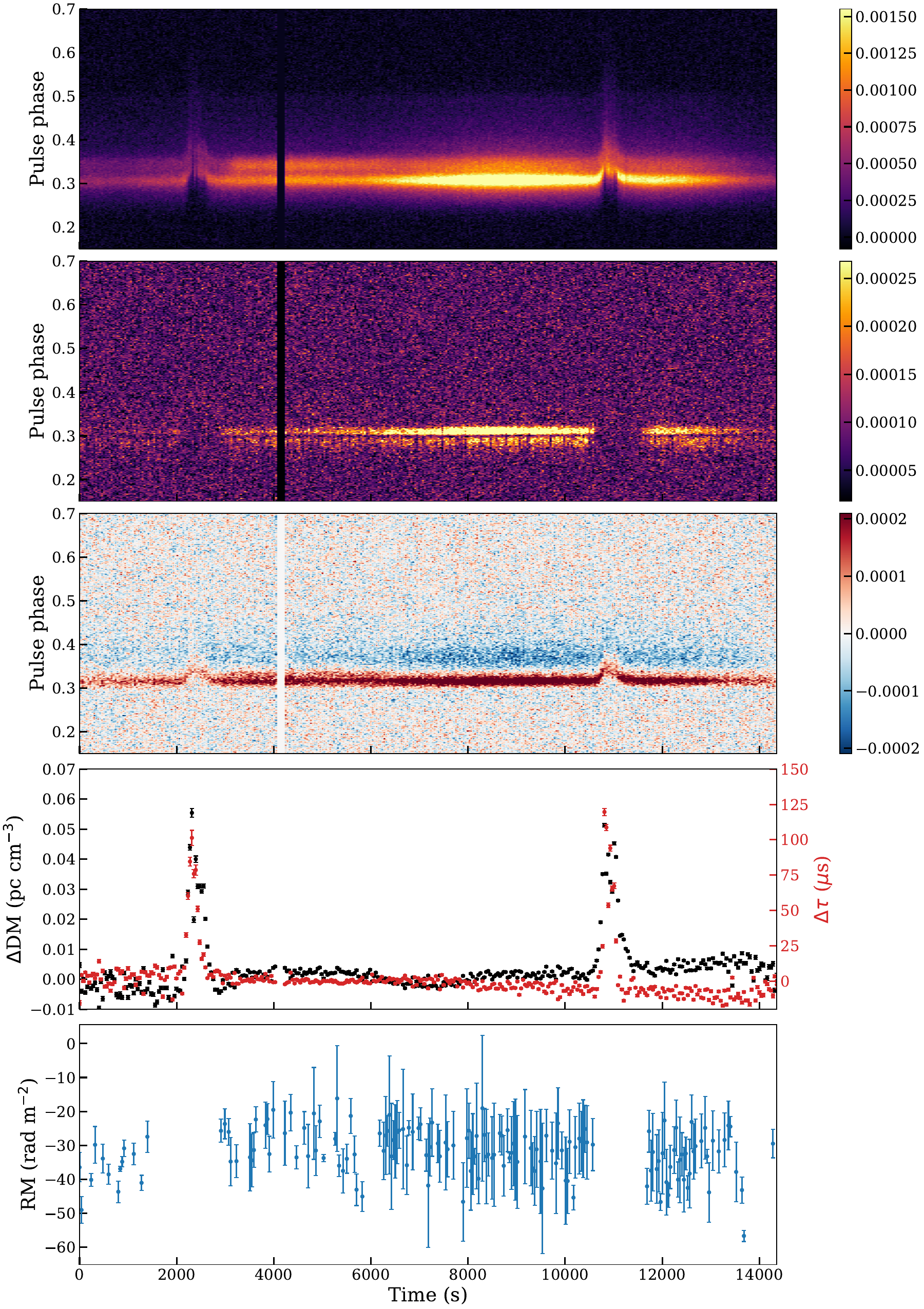}
    \caption{Evolution of total intensity (1st panel/top), linear polarisation intensity (2nd panel), circular polarisation intensity (3rd panel), DM and scattering variation (4th panel) and RM variation (5th panel) as a function of observation time for PSR J2051$-$0827, observed at L-band with 856 MHz bandwidth and 1284 MHz centre frequency. We observe two eclipses during our 4\,hrs observation.}
   \label{fig:phasetimeJ2051}
\end{figure*}

\begin{figure}
\centering
\includegraphics[width=0.95\linewidth]{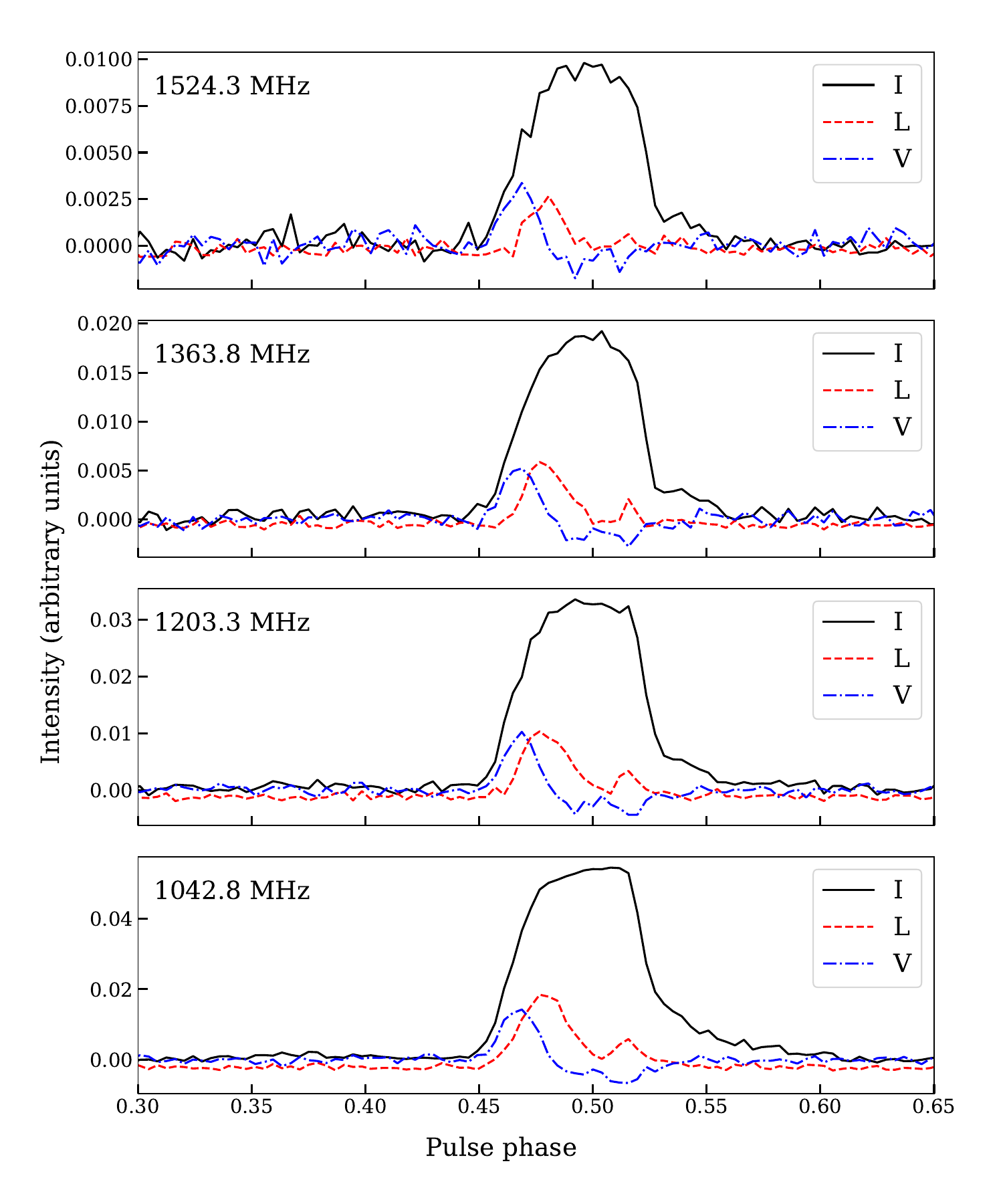}
\caption{The pulse profile of PSR J1748$-$2446A at different frequency channels for 320\,s of sub-integrated time, far away from the eclipse region. The solid black lines represent the total intensity, the red dashed lines represent linear polarisation and the blue dash-dotted lines represent circular polarisation. The effects of scattering are visible at the lower frequency channels. The linear and circular polarisations are equally strong at all frequency channels.}
\label{fig:profileJ1748}
\end{figure} 
\begin{figure*}
\includegraphics[width=0.75\linewidth]{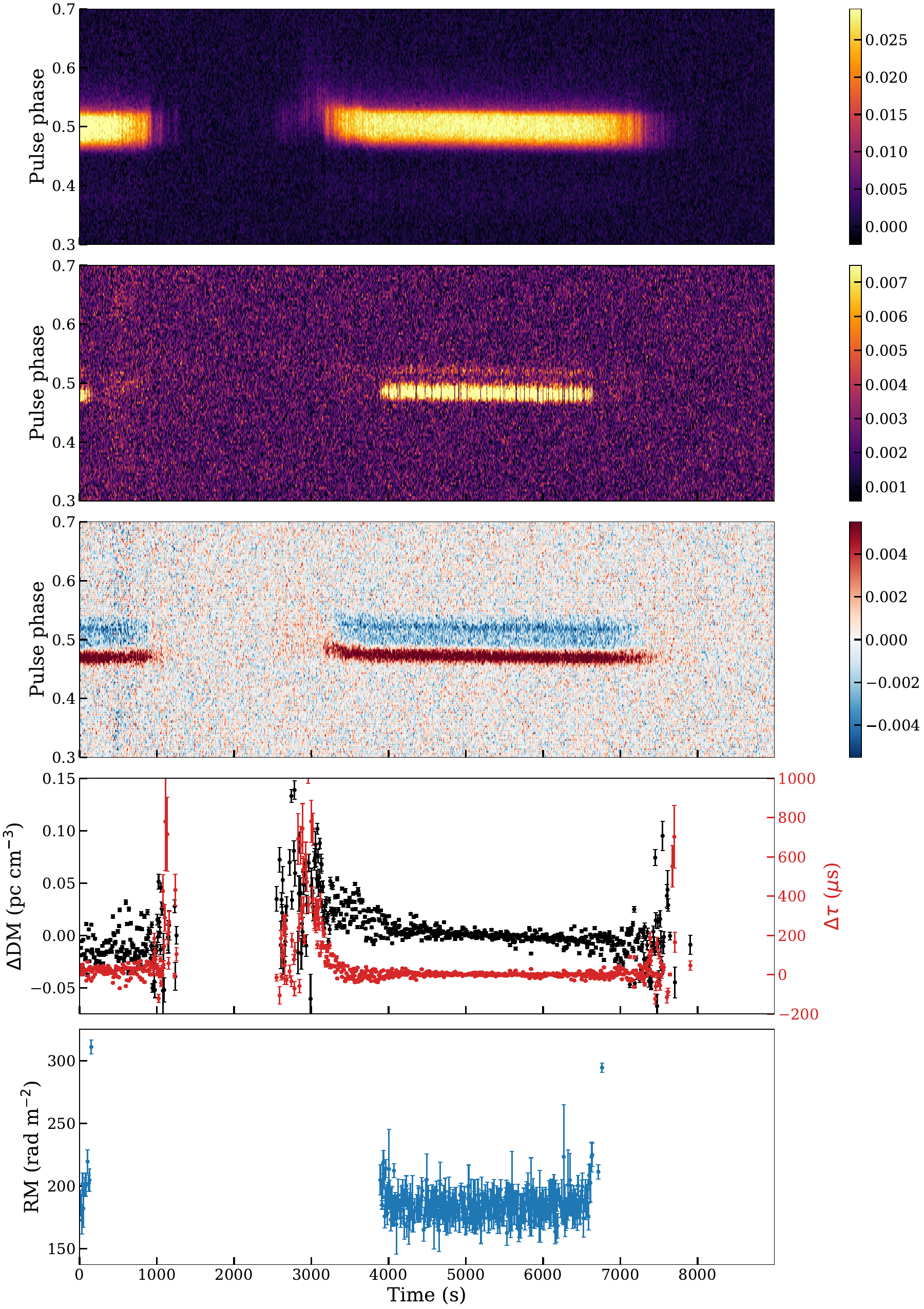}
\caption{Evolution of total intensity (1st panel/top), linear polarisation intensity (2nd panel), circular polarisation intensity (3rd panel), DM and scattering variation (4th panel) and RM variation (5th panel) as a function of observation time for PSR J1748$-$2444A, observed at L-band with 642 MHz bandwidth and 1283 MHz centre frequency. We observe two eclipses during our 2.5\,hrs observation. The pulsar is eclipsed for 30$\%$ of the orbit.}
\label{fig:phasetimeJ1748}
\end{figure*}
% Projects, sources  
Details of the data presented in this work, and planned for the upcoming papers in this series are given in Table~\ref{tab:pulsar_properties}. They were principally recorded as part of a MeerKAT Open Time proposal (OT4, PI: Li). We also include observations from the MeerTime Large Survey Proposal, specifically from the Globular Cluster programme (Proposal ID: SCI-20180516-MB-04).  
%Backends - fold mode and search mode.
The data were recorded using the Pulsar Timing User Supplied Equipment (PTUSE) backend in one or both of its two observing modes: fold-mode and search-mode (the latter also referred to as the single-pulse mode) \citep{Bailes_etal2020}. In the fold mode, the PTUSE backend applies an updated ephemeris, to produce data averaged to 8\,s time sub-integrations, folded using the known timing parameters and with a specified number of phase bins (typically 1024 bins) across the pulse period. During data acquisition, dispersive delays are removed coherently based on the ephemerides provided by the PI of the proposal before the observations. Search mode observations capture filterbank data without folding, and integrate the data thus preserving data as timeseries. This enables detailed post-processing analyses, including single pulse studies of the observed sources. This observing mode is particularly advantageous for globular cluster observations, where multiple pulsars can lie within the MeerKAT primary beam and can be recorded simultaneously. By subsequently folding the data using pulsar-specific timing ephemerides during the reduction stage, pulse profiles for multiple pulsars within the primary beam can be generated from the same observation.

PSR J1740$-$3052 and PSR J2051$-$0827 were recorded using both fold and search modes, where both the modes have full polarisation information, as a part of the aforementioned open time proposal. For these pulsars, we present results of the fold mode observations, recorded with the default 8~s per sub-integration, 1024 frequency channels across the observing band, and with 1024 phase bins across the pulse period. Additionally, for J1740$-$3052, we recorded the data in 4096 frequency channels, but found that 1024 channels were sufficient for the present study. PSR J1740$-$3052 and PSR J2051$-$0827 were observed in two different frequency bands centred at 2405 MHz (S1-band) with 875 MHz bandwidth, and centred at 816 MHz (UHF band) with 544 MHz bandwidth. We make use of Terzan~5 search mode data recorded as a part of the aforementioned MeerTime Large Survey Proposal. For these observations, the MeerKAT beam was centred on PSR J1748$-$2446O, approximately 35$^{\prime\prime}$ offset from  PSR J1748$-$2446A, with a native time resolution of 9.57 $\mu$s and 4096 frequency channels. We refolded these data using the best-known ephemerides of PSR J1748$-$2446A using \textsc{DSPSR} \citep{Straten_Bailes2010} to produce 8\,s~sub-integrated profiles with 768 frequency channels and 1024 phase bins. These observations were centred at 1284 MHz, with a bandwidth of 642 MHz. As these observations were conducted during the initial stages of the MeerTime programme, they have different bandwidths and channel numbers than more recent MeerTime observations. We have used the software \textsc{clfd} \citep{Morello_etal2019} to remove radio frequency interference (RFI) from our data.

All PTUSE L- and UHF-band observations (taken after April 2020 and August 2021 respectively) are by default polarisation calibrated by the observatory, producing calibrated Stokes products across frequency. Details of this calibration procedure are described in \citet{Serylak_etal2021}. We apply a final frontend correction to the L- and UHF-band Open Time data using \texttt{pac -XP} within \textsc{PSRCHIVE}. Polarisation calibration of the S1-band PTUSE observations\footnote{\url{https://skaafrica.atlassian.net/wiki/spaces/ESDKB/overview}} was not yet automated when the data for the present project were recorded. To calibrate these data, we applied polarisation solutions as computed ex post facto from the array calibration observations that preceded the science observation. These solutions, constructed as Jones matrices, are derived from calibration observations during the phasing up of the telescope and are stored in calibration files, which can be applied using the \textsc{PSRCHIVE} function and option, \texttt{pac -Q}. We also followed this procedure to calibrate the older (2019) Terzan~5 data that predate the automatic procedures.

We use the \texttt{dmfitter} package \citep{Lin_etal2021} to estimate the dispersion measures (DMs) and scattering timescales for each sub-integrated profiles produced after the reductions described above. This toolkit is designed to characterise frequency-dependent propagation effects that introduce arrival-time delays and pulse broadening, enabling precise measurements of both dispersion and scattering. The method employs template matching in the Fourier domain, following the framework developed by \citet{Penucci_etal2014}, to simultaneously model DM and scattering timescales. 

Rotation measure (RM) values were measured using the \texttt{rmfit} function within the \textsc{PSRCHIVE} \citep{Hota_etal2004, Straten_etal2012} software package. To assess the reliability of the RM measurements, we applied an additional validation procedure. For each RM estimation, we computed the linearly polarised intensity, $L = \sqrt{Q^{2} + U^{2}}$, and compared the signal-to-noise ratio (S/N) of the linear polarisation before and after applying the fitted RM correction. An RM measurement was considered reliable if the corrected data showed an improvement in the linear-polarisation S/N of at least 10\%. Measurements that did not produce a measurable increase in linear S/N were discarded. In addition, for orbital phases near eclipse, where propagation effects are strongest, we visually inspected the \texttt{rmfit} solutions to confirm that the derived RM values produced physically consistent polarisation behaviour and an appropriate alignment of the linear polarisation across frequency.
\begin{figure*}
\includegraphics[width=0.7\linewidth]{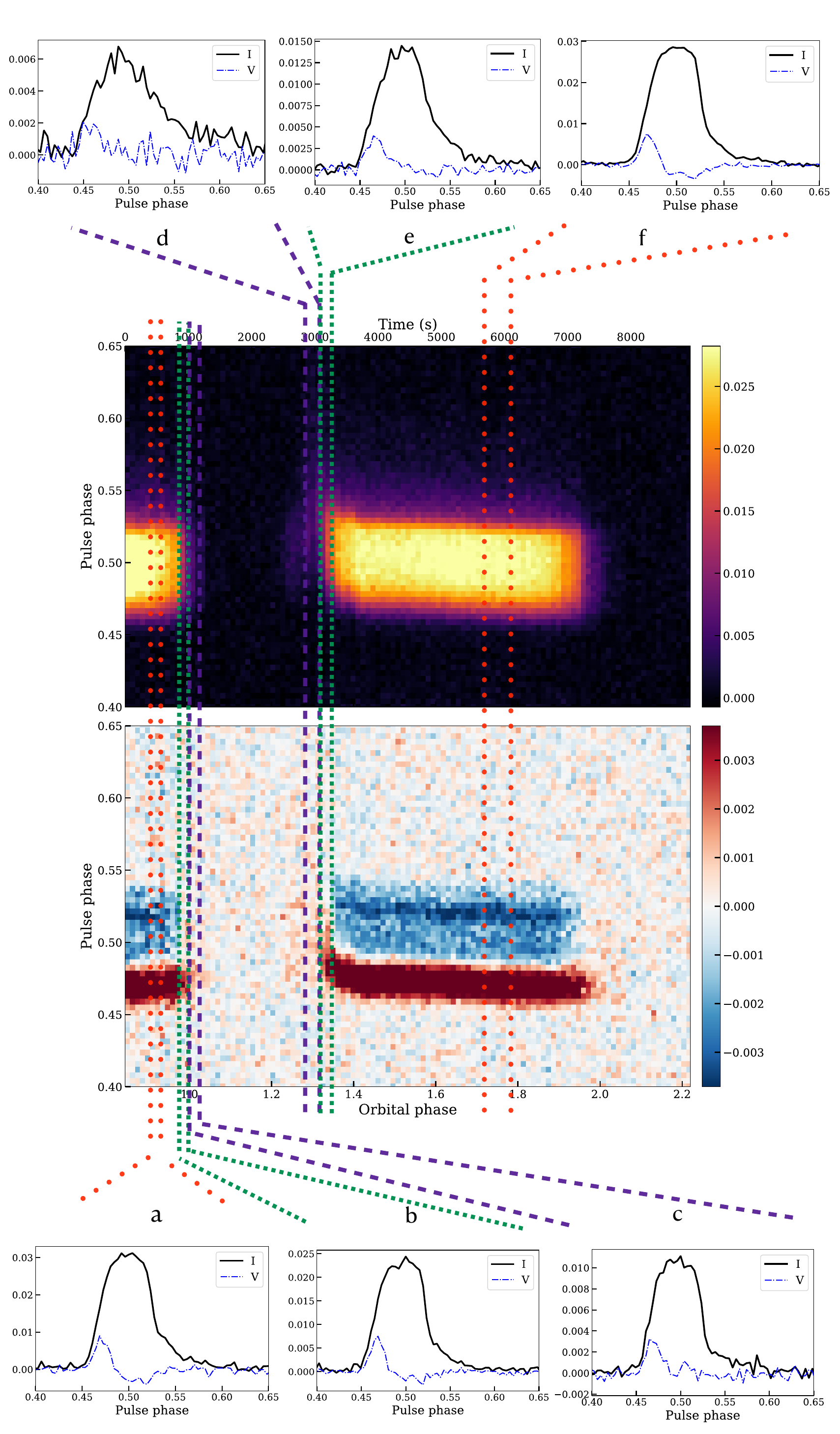}
\caption{Propagation effects seen in PSR J1748$-$2444A as the line of sight moves towards and away from the eclipse. Plots a and e represent the normal mode far away from the eclipse region, plots b, c, d and f represents the profile as the pulsar enters and exits the eclipse. The circular polarisation evolves and the second component seems to have a very small change in sign during these phases. This could be due Faraday Conversion or scattering or a combination of the two. This requires a detailed study and we defer it to our future publication.}
\label{fig:faradayconvJ1748}
\end{figure*}
\section{Results and Discussions}
\label{sec:results}
We present here the characteristics and results of the pulsar binaries obtained in our sample. Out of all the observations, we select one as a representative dataset and discuss our findings. 

\subsection{PSR J1740$-$3052}
PSR J1740$-$3052 is a young, long-period pulsar in a wide and highly eccentric ($e \sim 0.57$) binary system with a massive stellar companion \citep{Stairs_etal2001}, discovered in the Parkes Multibeam Survey \citep{Manchester_etal2001}. The pulsar has a spin period of $P \simeq 570$~ms and orbits its companion with a period of approximately $231$~days. An improved timing solution \citep{Madsen_etal2012} of this pulsar has enabled the measurement of the derivative of the projected semi-major axis, indicating that the orbital plane is significantly misaligned with the companion's spin axis. Such a tilt likely originated from asymmetries in the supernova explosion that formed the neutron star, implying a natal kick velocity of at least $\sim 50~\mathrm{km\,s^{-1}}$ \citep{Madsen_etal2012}. Additional timing residuals show quasi-periodic behaviour on a timescale of approximately $2.2\,P_{\rm B}$, the origin of which remains uncertain. Given the long orbital period, we do not expect to see radio eclipses at many orbital phases. Even near periastron passage we do not observe the system to eclipse. However, we do observe strong propagation effects near periastron passage, where the pulsar signal traverses the dense stellar wind of the companion. Long-term timing measurements have revealed orbital-phase-dependent variations in both pulse scattering and dispersion of the pulsar’s radio emission \citep{Madsen_etal2012}. These variations provide strong evidence that the pulsar signal propagates through a dense stellar wind originating from the companion star. 

PSR J1740$-$3052 was observed in two bands using MeerKAT. Here we present the results of one of the S1-band observations close to the periastron and the pulsar profile as a function of frequency for a 320~s sub-integrated profile is shown in Fig.~\ref{fig:profileJ1740}. We use 80\,s of sub-integrated profiles to estimate DM, RM and scattering time scales. This is a system with a very wide orbit and given that we do not observe eclipses we choose to add all the profiles to obtain a high S/N template. There is no scattering evident in the S1-band observations. A slow variation in RM is seen (see Fig.~\ref{fig:phasetimeJ1740}), with a maximum change of $\sim$ 200 rad\,m$^2$. Linear polarisation is not seen in some of the sub-integrations despite having high S/N total intensity profiles. This effect is prominently seen in long term observations of this pulsar in L-band (to be discussed in detail in a future paper). The cause of this depolarisation is not fully known but it could be because of the magneto-ionic environment of the companion. We observed a change in RM as high as 117 rad\,m$^{-2}$ with respect to the median and a change in DM not more than 0.29 pc\,cm$^{-3}$ giving a very low magnetic limit of $\langle$B$_{\parallel}\rangle$ = 1.2 $\mu$G ($\Delta $RM$/117\,$rad$ \ $m$^{-2} $)/($\Delta $\rm{DM}$/0.29$\,pc cm$^{-3}$) $>$ 0.5\,mG. We observe a slow change in the value of RM ($\Delta_{\rm{RM}} \sim$ 150 rad m$^{-2}$) on a time scale of $\sim$1500\,s close to the epoch of periastron. Based on approximate binary parameters of the system, we can compute a rough estimate of the the magnetic size/spatial scale associated with the observed slow varying RM changes. Assuming a binary separation of around 0.75 AU and the companion mass of close to 11 to 16 M$_\odot$ \citep{Stairs_etal2001, Bassa_etal2011}, and the eccentricity to be around 0.6, we can estimate the relative velocity near periastron to be $\sim$250 km s$^{-1}$, which is expected to be larger than the typical stellar wind velocity of 30 km s$^{-1}$ \citep{Dupree1986, Stairs_etal2001}. The gradual change in RM ($\Delta_{\rm{RM}} \sim$ 150 rad m$^{-2}$) would suggest a magnetic spatial scale of $\sim 4 \times 10^{5}$\,km (0.003 AU). This value is an order of magnitude larger than the light cylinder radius of this pulsar ($\sim 0.3 \times 10^{5}$\,km) and therefore the observed RM structure is unlikely related to the stripped wind of the pulsar, and more likely related to the turbulent scale in the companion wind or near the interphase, where the pulsar wind and the companion star’s stellar wind meet or interact. 

%This distance is a few factors more than the light cylinder radius of this system, suggesting the observed changes arise due to the interaction of the stellar wind with the magnetosphere.

\subsection{PSR J2051$-$0827}
PSR J2051$-$0827 \citep{Stappers_etal1996} is a millisecond black widow pulsar binary and one of the most studied systems of its class. It was discovered in the Parkes Southern Pulsar Survey \citep{Manchester_etal1996}, and is a recycled millisecond pulsar with a spin period of approximately $4.5$~ms, orbiting a very low-mass companion in a compact binary with an orbital period of $\sim 2.4$~hr. The companion mass is estimated to be $\sim 0.02$--$0.05~M_{\odot}$, consistent with a semi-degenerate object that is being ablated by the pulsar wind. The system exhibits pronounced radio eclipses over $\sim10\%$ of the orbit at a wide range of radio frequencies \citep{Stappers_etal2001, Polzin_etal2019, Lin_etal2021, Wang_etal2023}. These eclipses are strongly frequency dependent and are accompanied by orbital-phase-dependent variations in pulse arrival times and DM. Such behaviour indicates the presence of dense plasma local to the binary, most likely originating from ionised material driven off the companion by irradiation from the pulsar's relativistic wind. Measurements of Faraday rotation, depolarisation, and pulse profile evolution imply the presence of magnetised plasma within the binary system, with an estimated line-of-sight magnetic field strength in the eclipse medium of $\sim$0.1\,G \citep{Wang_etal2023}. Long term timing observations of PSR J2051$-$0827 have revealed secular and quasi-stochastic changes in the orbital period and projected semi-major axis \citep{Shaifullah_etal2016}, commonly interpreted as arising from variations in the companion's gravitational quadrupole moment.  
%The combination of its short orbital period, strong and variable eclipses, and pronounced magneto-ionic signatures makes PSR J2051$-$0827 a key target for probing intra-binary plasma density, magnetic field structure, and their evolution with orbital phase.

We present here results from one of the observations of PSR J2051$-$0827 in L-band which spans $\sim$4\,hrs of on-source time, capturing two eclipses during the observation. The pulse profile obtained from 320~s of sub-integrated time is shown in the left panel of Fig.~\ref{fig:profileJ2051} at four different frequency channels. This pulsar has a DM value of around 20.2 pc cm$^{-3}$ and we observe the total intensity profile to evolve as a function of frequency showing scintillation. It is also seen that the linear and circular polarisation intensities vary as a function of observing frequency. The waterfall plots (shown in the right panel of Fig.~\ref{fig:profileJ2051}) of Stokes Q and U exhibit multi-component features, and a clear evolution of  these components with frequency. We see e.g. a stronger leading component in both Q and U emerge at lower frequencies. This is similarly seen plotted as the linear polarisation profile in the left-hand side panel. While the cause of this is not fully known, it is important to note that the linear polarisation profiles have been seen to vary as a function of both frequency and time in a magnetar \citep{Lower_etal2024} providing a unique probe of the relativistic, magnetized plasma within the near-field environment of these ultra-magnetic neutron stars. The pulsar undergoes two eclipses during our observation as seen in Fig.~\ref{fig:phasetimeJ2051}. The evolution of total intensity, linear and circular polarisation across the eclipse can been seen in the first three panels of this figure. A characteristic feature of this system is that the pulsar is not fully eclipsed by its companion, as evidenced by the continued detection of pulsar flux, albeit at a reduced level, during the eclipse phase. We observed linear depolarisation during this phase. Using 40\,s sub-integrated profiles we estimate the DM, RMs and scattering timescales for this source. This choice was made to achieve a high S/N for reliable RM estimation. A template was created by summing  all the profiles in the region where no eclipses were observed (between 4000~s to 9500~s of observation). The changes in the DM and scattering timescales were estimated with respect to this template. The variation of DM, scattering timescales and RM is shown in Fig~\ref{fig:profileJ2051}. \cite{Wang_etal2023} observed gradual changes in RM from 60 rad m$^{-2}$ to 28.7 rad m$^{-2}$ and RM gradually changing back to the template value when the line of sight moved away from the eclipse. While we do not see such apparent changes in our observations, the eclipse of linear polarization does imply a rapidly changing RM.

\subsection{PSR J1748$-$2446A}
PSR J1748$-$2446A (or Terzan~5A, abbreviated as Ter5A), is an eclipsing millisecond pulsar, now classified as redback system (previously as a black widow, \citealt{Roberts2013}) with a rotational period of 11.56 ms located in the dense stellar environment of the globular cluster Terzan~5. It was first observed using the Parkes Radio telescope \citep{Lyne_etal1990} and has a short orbital period of 1.8\,hrs, indicating that the two objects are extremely close together with a separation of $\sim0.85\,R_{\odot}$. The companion mass is relatively low, estimated to be at least 0.085 M$_{\odot}$, indicating that the companion is likely a low-mass star.
The eclipse duration in this system is both time variable and strongly dependent on observing frequency \citep{You_etal2018, Bilous_etal2019, Li_etal2023}. At $\sim$1.4~GHz, the eclipse typically spans more than one-third of the orbital period. The eclipses here are believed to be caused by ionised material extending well beyond the companion, most likely an ablated wind driven by the pulsar irradiation. Multi-frequency observations between 0.8 and 1.66~GHz further showed that the eclipse duration decreases with observing frequency, following an approximate dependence of $\nu^{-0.63 \pm 0.18}$ \citep{Nice_etal1990}. PSR J1748$-$2446A has been observed to exhibit both extreme plasma lensing \citep{Bilous_etal2019} and Faraday Conversion phenomena \citep{Li_etal2023}. Since one of the requirements for Faraday Conversion to occur is the presence of large magnetic fields \citep{Li_etal2023}, the observation of this phenomena in this system at $\sim$2~GHz place constraints on the magnetic field strength within the companion’s magnetosphere, providing B $>10$\,G. Long-term timing observations spanning 34~yr reveal that the system exhibits unusually large spin irregularities compared to other globular cluster pulsars, reflecting significant timing noise likely associated with the interactions between the pulsar wind and material from the companion \citep{Rosenthal_etal2025}. Despite this variability, no strong correlation is observed between spin and orbital changes. The measured orbital period contraction is broadly consistent with general relativistic decay, with additional contributions from redback-driven orbital variability constrained to be relatively small. 

We have used the L-band search-mode Meertime observations of the Terzan~5 Globular Cluster, with full Stokes information and high time resolution.  We present results for one of these data sets, folded on PSR J1748$-$2446A. The pulsar profile as a function of frequency for a 320~s sub-integrated profile is shown in Fig.~\ref{fig:profileJ1748}. The pulsar is scattered at lower frequencies. Our observation spans more than one orbit and the pulsar undergoes almost two eclipses during this time, as shown in Fig.~\ref{fig:phasetimeJ1748}. We observed that the pulsar remains in eclipse for a large fraction of the orbital period ($30\%$). We use 8\,s sub-integrated profiles to estimate DMs, RMs and scattering time scales. A template was generated by adding all the profiles in the region away from the eclipse (4100~s to 6500~s). Changes in DMs and scattering time scales were estimated with respect to this template. The variation in DM, RM and scattering timescales is shown in Fig.~\ref{fig:phasetimeJ1748}. The maximum RM change seen for this pulsar is around 127 rad $m^{-2}$ and a corresponding change in DM not more than 0.01 pc $cm^{-3}$, where the pulsar is about to enter the eclipse. This provides a lower limit of the average intervening magnetic field of $\langle$B$_{\parallel}\rangle$ = 1.2 $\mu$G ($\Delta $RM$/ 127$\,rad\ m$^{-2} $)/($\Delta$DM$/ 0.01 $\,pc cm$^{-3}$) $>$ 15\,mG. The excess DM of about 0.15 pc cm$^{-3}$ during the eclipse suggest an RM change of 2286 rad m$^{-2}$ given the limits of magnetic field, but we do not observed such changes here. This could be partly due to rapid RM changes during the integration time and partly due to the effects of scattering. \cite{Li_etal2023} observed a sign reversal of the circular polarisation during the eclipse (when the pulsar is behind the companion) indicating Faraday Conversion caused by the plasma and magnetic field of the companion. A strong sign flip is not observed in our polarimetric data; however, there is a subtle indication of a reversal in the sign of the circular polarisation. This could be due to Faraday Conversion or scattering. A detailed view of the circular polarisation intensities are presented in Fig.\ref{fig:faradayconvJ1748}. The panels (a) and (f) represent the normal state, far away from the eclipse region, whereas panels (b), (c), (d) and (e) show the profiles as the pulsar enters and exits the eclipse. There are hints of sign reversal in the circular polarisation during the ingress and egress. Ongoing investigations will confirm whether these changes are due to scattering or Faraday Conversion.

\section{Summary and Future Work}
\label{sec:summary}
Binary pulsars in which we observe strong imprints of magneto-ionic propagation effects, allow us to constraint the intra-binary plasma model and the companion magnetic fields. We have presented here the first results from MeerKAT Open Time study aimed at understanding the magneto-active environments of pulsar binary systems through broadband polarimetric observations. The primary goal of this work is to present the observational properties of our sample of binaries and to provide a foundation for detailed studies of plasma lensing and eclipse mechanisms in subsequent papers of this series. Our sample consists of three systems: the high-mass binary PSR~J1740$-$3052, the black widow PSR~J2051$-$0827 and the redback PSR~J1748$-$2446A (Terzan~5A). Using high-sensitivity MeerKAT observations, we characterised their polarisation behaviour, and temporal evolution of dispersion measure (DM), rotation measure (RM), and scattering time scales.

The wide orbit binary system PSR~J1740$-$3052 shows no eclipses but displays gradual RM variability with minimal scattering at S1-band frequencies. These observations are consistent with propagation through an extended stellar wind rather than a confined intra-binary eclipse medium, demonstrating how pulsar observations probe magneto-ionic environments across a wide range of spatial scales. The two eclipsing millisecond pulsars exhibit strong orbital-phase-dependent propagation effects. In both PSR~J2051$-$0827 and PSR~J1748$-$2446A, we observe excess dispersion but no significant RM variability near eclipse. We see evolution of linear polarisation in PSR~J2051$-$0827. In PSR~J1748$-$2446A, the evolution of circular polarisation during eclipse ingress and egress provides evidence for propagation effects in this system. The measurements presented here establish baseline properties for the observed systems and motivate further investigation using higher time-resolution search-mode data. Future work in this series will focus on identifying plasma lensing events, constraining eclipse mechanism and geometries, and understanding depolarisation in high mass binaries.

\section*{Acknowledgements}
The MeerKAT telescope is operated by the South African Radio Astronomy Observatory (SARAO), which is a facility of the National Research Foundation, an agency of the Department of Science and Innovation. SARAO acknowledges the ongoing advice and calibration of GPS systems by the National Metrology Institute of South Africa (NMISA) and the time space reference systems department of the Paris Observatory. PTUSE was developed with support from the Australian SKA Oﬃce and Swinburne University of Technology. This work made use of the OzSTAR national HPC facility at Swinburne University of Technology. MeerTime data are housed on the OzSTAR supercomputer. The OzSTAR program receives funding in part from the Astronomy National Collaborative Research Infrastructure Strategy (NCRIS) allocation provided by the Australian Government. We acknowledge the use of the ilifu cloud computing facility – www.ilifu.ac.za, a partnership between the University of Cape Town, the University of the Western Cape, Stellenbosch University, Sol Plaatje University and the Cape Peninsula University of Technology. The ilifu facility is supported by contributions from the Inter-University Institute for Data Intensive Astronomy (IDIA – a partnership between the University of Cape Town, the University of Pretoria and the University of the Western Cape), the Computational Biology division at UCT and the Data Intensive Research Initiative of South Africa (DIRISA). JS, MG and AW acknowledge the support from the University of Cape Town Vice Chancellor’s Future Leaders 2030 Awards programme and the South African Research Chairs Initiative of the Department of Science and Technology and the National Research Foundation.

%%%%%%%%%%%%%%%%%%%%%%%%%%%%%%%%%%%%%%%%%%%%%%%%%%
\section*{Data Availability}
The data used in this paper will be provided on request.

%%%%%%%%%%%%%%%%%%%% REFERENCES %%%%%%%%%%%%%%%%%%

% The best way to enter references is to use BibTeX:

\bibliographystyle{mnras}
\bibliography{reference} % if your bibtex file is called example.bib

@INPROCEEDINGS{Roberts2013,
       author = {{Roberts}, Mallory S.~E.},
        title = "{Surrounded by spiders! New black widows and redbacks in the Galactic field}",
    booktitle = {Neutron Stars and Pulsars: Challenges and Opportunities after 80 years},
         year = 2013,
       editor = {{van Leeuwen}, Joeri},
       series = {IAU Symposium},
       volume = {291},
        month = mar,
        pages = {127-132},
          doi = {10.1017/S174392131202337X},
       adsurl = {https://ui.adsabs.harvard.edu/abs/2013IAUS..291..127R}
}

@ARTICLE{Main_etal2018,
       author = {{Main}, Robert and {Yang}, I.-Sheng and {Chan}, Victor and {Li}, Dongzi and {Lin}, Fang Xi and {Mahajan}, Nikhil and {Pen}, Ue-Li and {Vanderlinde}, Keith and {van Kerkwijk}, Marten H.},
        title = "{Pulsar emission amplified and resolved by plasma lensing in an eclipsing binary}",
      journal = {\nat},
         year = 2018,
        month = may,
       volume = {557},
       number = {7706},
        pages = {522-525},
          doi = {10.1038/s41586-018-0133-z},
archivePrefix = {arXiv},
       eprint = {1805.09348},
 primaryClass = {astro-ph.HE},
       adsurl = {https://ui.adsabs.harvard.edu/abs/2018Natur.557..522M}
}

@ARTICLE{Bilous_etal2019,
       author = {{Bilous}, A.~V. and {Ransom}, S.~M. and {Demorest}, P.},
        title = "{Unusually Bright Single Pulses from the Binary Pulsar B1744-24A: A Case of Strong Lensing?}",
      journal = {\apj},
         year = 2019,
        month = jun,
       volume = {877},
       number = {2},
          eid = {125},
        pages = {125},
          doi = {10.3847/1538-4357/ab16dd},
archivePrefix = {arXiv},
       eprint = {1811.05766},
 primaryClass = {astro-ph.HE},
       adsurl = {https://ui.adsabs.harvard.edu/abs/2019ApJ...877..125B}
}

@ARTICLE{Johnston_etal2005,
       author = {{Johnston}, Simon and {Ball}, Lewis and {Wang}, N. and {Manchester}, R.~N.},
        title = "{Radio observations of PSR B1259-63 through the 2004 periastron passage}",
      journal = {\mnras},
         year = 2005,
        month = apr,
       volume = {358},
       number = {3},
        pages = {1069-1075},
          doi = {10.1111/j.1365-2966.2005.08854.x},
archivePrefix = {arXiv},
       eprint = {astro-ph/0501660},
 primaryClass = {astro-ph},
       adsurl = {https://ui.adsabs.harvard.edu/abs/2005MNRAS.358.1069J}
}

@ARTICLE{Li_etal2023,
       author = {{Li}, Dongzi and {Bilous}, Anna and {Ransom}, Scott and {Main}, Robert and {Yang}, Yuan-Pei},
        title = "{A highly magnetized environment in a pulsar binary system}",
      journal = {\nat},
         year = 2023,
        month = jun,
       volume = {618},
       number = {7965},
        pages = {484-488},
          doi = {10.1038/s41586-023-05983-z},
archivePrefix = {arXiv},
       eprint = {2205.07917},
 primaryClass = {astro-ph.HE},
       adsurl = {https://ui.adsabs.harvard.edu/abs/2023Natur.618..484L}
}

@INPROCEEDINGS{Jonas_etal2016,
       author = {{Jonas}, J. and {MeerKAT Team}},
        title = "{The MeerKAT Radio Telescope}",
    booktitle = {MeerKAT Science: On the Pathway to the SKA},
         year = 2016,
        month = jan,
          eid = {1},
        pages = {1},
          doi = {10.22323/1.277.0001},
       adsurl = {https://ui.adsabs.harvard.edu/abs/2016mks..confE...1J}
}

@ARTICLE{Bailes_etal2020,
       author = {{Bailes}, M. and {Jameson}, A. and {Abbate}, F. and {Barr}, E.~D. and {Bhat}, N.~D.~R. and {Bondonneau}, L. and {Burgay}, M. and {Buchner}, S.~J. and {Camilo}, F. and {Champion}, D.~J. and {Cognard}, I. and {Demorest}, P.~B. and {Freire}, P.~C.~C. and {Gautam}, T. and {Geyer}, M. and {Griessmeier}, J.-M. and {Guillemot}, L. and {Hu}, H. and {Jankowski}, F. and {Johnston}, S. and {Karastergiou}, A. and {Karuppusamy}, R. and {Kaur}, D. and {Keith}, M.~J. and {Kramer}, M. and {van Leeuwen}, J. and {Lower}, M.~E. and {Maan}, Y. and {McLaughlin}, M.~A. and {Meyers}, B.~W. and {Os{\l}owski}, S. and {Oswald}, L.~S. and {Parthasarathy}, A. and {Pennucci}, T. and {Posselt}, B. and {Possenti}, A. and {Ransom}, S.~M. and {Reardon}, D.~J. and {Ridolfi}, A. and {Schollar}, C.~T.~G. and {Serylak}, M. and {Shaifullah}, G. and {Shamohammadi}, M. and {Shannon}, R.~M. and {Sobey}, C. and {Song}, X. and {Spiewak}, R. and {Stairs}, I.~H. and {Stappers}, B.~W. and {van Straten}, W. and {Szary}, A. and {Theureau}, G. and {Venkatraman Krishnan}, V. and {Weltevrede}, P. and {Wex}, N. and {Abbott}, T.~D. and {Adams}, G.~B. and {Burger}, J.~P. and {Gamatham}, R.~R.~G. and {Gouws}, M. and {Horn}, D.~M. and {Hugo}, B. and {Joubert}, A.~F. and {Manley}, J.~R. and {McAlpine}, K. and {Passmoor}, S.~S. and {Peens-Hough}, A. and {Ramudzuli}, Z.~R. and {Rust}, A. and {Salie}, S. and {Schwardt}, L.~C. and {Siebrits}, R. and {Van Tonder}, G. and {Van Tonder}, V. and {Welz}, M.~G.},
        title = "{The MeerKAT telescope as a pulsar facility: System verification and early science results from MeerTime}",
      journal = {\pasa},
         year = 2020,
        month = jul,
       volume = {37},
          eid = {e028},
        pages = {e028},
          doi = {10.1017/pasa.2020.19},
archivePrefix = {arXiv},
       eprint = {2005.14366},
 primaryClass = {astro-ph.IM},
       adsurl = {https://ui.adsabs.harvard.edu/abs/2020PASA...37...28B}
}

@ARTICLE{Stappers_etal1996,
       author = {{Stappers}, B.~W. and {Bailes}, M. and {Lyne}, A.~G. and {Manchester}, R.~N. and {D'Amico}, N. and {Tauris}, T.~M. and {Lorimer}, D.~R. and {Johnston}, S. and {Sandhu}, J.~S.},
        title = "{Probing the Eclipse Region of a Binary Millisecond Pulsar}",
      journal = {\apjl},
         year = 1996,
        month = jul,
       volume = {465},
        pages = {L119},
          doi = {10.1086/310148},
       adsurl = {https://ui.adsabs.harvard.edu/abs/1996ApJ...465L.119S}
}

@ARTICLE{Manchester_etal1996,
       author = {{Manchester}, R.~N. and {Lyne}, A.~G. and {D'Amico}, N. and {Bailes}, M. and {Johnston}, S. and {Lorimer}, D.~R. and {Harrison}, P.~A. and {Nicastro}, L. and {Bell}, J.~F.},
        title = "{The Parkes Southern Pulsar Survey. I. Observing and data analysis systems and initial results.}",
      journal = {\mnras},
         year = 1996,
        month = apr,
       volume = {279},
       number = {4},
        pages = {1235-1250},
          doi = {10.1093/mnras/279.4.1235},
       adsurl = {https://ui.adsabs.harvard.edu/abs/1996MNRAS.279.1235M}
}

@ARTICLE{Stappers_etal2001,
       author = {{Stappers}, B.~W. and {Bailes}, M. and {Lyne}, A.~G. and {Camilo}, F. and {Manchester}, R.~N. and {Sandhu}, J.~S. and {Toscano}, M. and {Bell}, J.~F.},
        title = "{The nature of the PSR J2051-0827 eclipses}",
      journal = {\mnras},
         year = 2001,
        month = mar,
       volume = {321},
       number = {3},
        pages = {576-584},
          doi = {10.1046/j.1365-8711.2001.04074.x},
       adsurl = {https://ui.adsabs.harvard.edu/abs/2001MNRAS.321..576S}
}

@ARTICLE{Polzin_etal2019,
       author = {{Polzin}, E.~J. and {Breton}, R.~P. and {Stappers}, B.~W. and {Bhattacharyya}, B. and {Janssen}, G.~H. and {Os{\l}owski}, S. and {Roberts}, M.~S.~E. and {Sobey}, C.},
        title = "{Long-term variability of a black widow's eclipses - A decade of PSR J2051-0827}",
      journal = {\mnras},
         year = 2019,
        month = nov,
       volume = {490},
       number = {1},
        pages = {889-908},
          doi = {10.1093/mnras/stz2579},
archivePrefix = {arXiv},
       eprint = {1909.06130},
 primaryClass = {astro-ph.HE},
       adsurl = {https://ui.adsabs.harvard.edu/abs/2019MNRAS.490..889P}
}

@ARTICLE{Wang_etal2023,
       author = {{Wang}, S.~Q. and {Wang}, J.~B. and {Li}, D.~Z. and {Yao}, J.~M. and {Manchester}, R.~N. and {Hobbs}, G. and {Wang}, N. and {Dai}, S. and {Xu}, H. and {Luo}, R. and {Feng}, Y. and {Wang}, W.~Y. and {Li}, D. and {Yu}, Y.~W. and {Du}, Z.~X. and {Niu}, C.~H. and {Zhang}, S.~B. and {Zhang}, C.~M.},
        title = "{Change of Rotation Measure during the Eclipse of a Black Widow PSR J2051-0827}",
      journal = {\apj},
         year = 2023,
        month = sep,
       volume = {955},
       number = {1},
          eid = {36},
        pages = {36},
          doi = {10.3847/1538-4357/acea81},
archivePrefix = {arXiv},
       eprint = {2307.13198},
 primaryClass = {astro-ph.HE},
       adsurl = {https://ui.adsabs.harvard.edu/abs/2023ApJ...955...36W}
}

@ARTICLE{Shaifullah_etal2016,
       author = {{Shaifullah}, G. and {Verbiest}, J.~P.~W. and {Freire}, P.~C.~C. and {Tauris}, T.~M. and {Wex}, N. and {Os{\l}owski}, S. and {Stappers}, B.~W. and {Bassa}, C.~G. and {Caballero}, R.~N. and {Champion}, D.~J. and {Cognard}, I. and {Desvignes}, G. and {Graikou}, E. and {Guillemot}, L. and {Janssen}, G.~H. and {Jessner}, A. and {Jordan}, C. and {Karuppusamy}, R. and {Kramer}, M. and {Lazaridis}, K. and {Lazarus}, P. and {Lyne}, A.~G. and {McKee}, J.~W. and {Perrodin}, D. and {Possenti}, A. and {Tiburzi}, C.},
        title = "{21 year timing of the black-widow pulsar J2051-0827}",
      journal = {\mnras},
         year = 2016,
        month = oct,
       volume = {462},
       number = {1},
        pages = {1029-1038},
          doi = {10.1093/mnras/stw1737},
archivePrefix = {arXiv},
       eprint = {1607.04167},
 primaryClass = {astro-ph.HE},
       adsurl = {https://ui.adsabs.harvard.edu/abs/2016MNRAS.462.1029S}
}

@ARTICLE{You_etal2018,
       author = {{You}, X.~P. and {Manchester}, R.~N. and {Coles}, W.~A. and {Hobbs}, G.~B. and {Shannon}, R.},
        title = "{Polarimetry of the Eclipsing Pulsar PSR J1748-2446A}",
      journal = {\apj},
         year = 2018,
        month = nov,
       volume = {867},
       number = {1},
          eid = {22},
        pages = {22},
          doi = {10.3847/1538-4357/aadee0},
archivePrefix = {arXiv},
       eprint = {1809.01309},
 primaryClass = {astro-ph.HE},
       adsurl = {https://ui.adsabs.harvard.edu/abs/2018ApJ...867...22Y}
}

@software{Straten_Bailes2010,
       author = {{van Straten}, W. and {Bailes}, M.},
        title = "{DSPSR: Digital Signal Processing Software for Pulsar Astronomy}",
 howpublished = {Astrophysics Source Code Library, record ascl:1010.006},
         year = 2010,
        month = oct,
          eid = {ascl:1010.006},
archivePrefix = {ascl},
       eprint = {1010.006},
       adsurl = {https://ui.adsabs.harvard.edu/abs/2010ascl.soft10006V}
}

@ARTICLE{Hota_etal2004,
       author = {{Hotan}, A.~W. and {van Straten}, W. and {Manchester}, R.~N.},
        title = "{PSRCHIVE and PSRFITS: An Open Approach to Radio Pulsar Data Storage and Analysis}",
      journal = {\pasa},
         year = 2004,
        month = jan,
       volume = {21},
       number = {3},
        pages = {302-309},
          doi = {10.1071/AS04022},
archivePrefix = {arXiv},
       eprint = {astro-ph/0404549},
 primaryClass = {astro-ph},
       adsurl = {https://ui.adsabs.harvard.edu/abs/2004PASA...21..302H}
}

@ARTICLE{Morello_etal2019,
       author = {{Morello}, V. and {Barr}, E.~D. and {Cooper}, S. and {Bailes}, M. and {Bates}, S. and {Bhat}, N.~D.~R. and {Burgay}, M. and {Burke-Spolaor}, S. and {Cameron}, A.~D. and {Champion}, D.~J. and {Eatough}, R.~P. and {Flynn}, C.~M.~L. and {Jameson}, A. and {Johnston}, S. and {Keith}, M.~J. and {Keane}, E.~F. and {Kramer}, M. and {Levin}, L. and {Ng}, C. and {Petroff}, E. and {Possenti}, A. and {Stappers}, B.~W. and {van Straten}, W. and {Tiburzi}, C.},
        title = "{The High Time Resolution Universe survey - XIV. Discovery of 23 pulsars through GPU-accelerated reprocessing}",
      journal = {\mnras},
         year = 2019,
        month = mar,
       volume = {483},
       number = {3},
        pages = {3673-3685},
          doi = {10.1093/mnras/sty3328},
archivePrefix = {arXiv},
       eprint = {1811.04929},
 primaryClass = {astro-ph.IM},
       adsurl = {https://ui.adsabs.harvard.edu/abs/2019MNRAS.483.3673M}
}

@ARTICLE{Serylak_etal2021,
       author = {{Serylak}, M. and {Johnston}, S. and {Kramer}, M. and {Buchner}, S. and {Karastergiou}, A. and {Keith}, M.~J. and {Parthasarathy}, A. and {Weltevrede}, P. and {Bailes}, M. and {Barr}, E.~D. and {Camilo}, F. and {Geyer}, M. and {Hugo}, B.~V. and {Jameson}, A. and {Reardon}, D.~J. and {Shannon}, R.~M. and {Spiewak}, R. and {van Straten}, W. and {Venkatraman Krishnan}, V.},
        title = "{The thousand-pulsar-array programme on MeerKAT IV: Polarization properties of young, energetic pulsars}",
      journal = {\mnras},
         year = 2021,
        month = aug,
       volume = {505},
       number = {3},
        pages = {4483-4495},
          doi = {10.1093/mnras/staa2811},
archivePrefix = {arXiv},
       eprint = {2009.05797},
 primaryClass = {astro-ph.HE},
       adsurl = {https://ui.adsabs.harvard.edu/abs/2021MNRAS.505.4483S}
}

@ARTICLE{Straten_etal2012,
       author = {{van Straten}, Willem and {Demorest}, Paul and {Oslowski}, Stefan},
        title = "{Pulsar Data Analysis with PSRCHIVE}",
      journal = {Astronomical Research and Technology},
         year = 2012,
        month = jul,
       volume = {9},
       number = {3},
        pages = {237-256},
          doi = {10.48550/arXiv.1205.6276},
archivePrefix = {arXiv},
       eprint = {1205.6276},
 primaryClass = {astro-ph.IM},
       adsurl = {https://ui.adsabs.harvard.edu/abs/2012AR&T....9..237V}
}

@ARTICLE{Lin_etal2021,
       author = {{Lin}, F.~X. and {Main}, R.~A. and {Verbiest}, J.~P.~W. and {Kramer}, M. and {Shaifullah}, G.},
        title = "{Discovery and modelling of broad-scale plasma lensing in black-widow pulsar J2051 - 0827}",
      journal = {\mnras},
         year = 2021,
        month = sep,
       volume = {506},
       number = {2},
        pages = {2824-2835},
          doi = {10.1093/mnras/stab1811},
archivePrefix = {arXiv},
       eprint = {2106.12359},
 primaryClass = {astro-ph.HE},
       adsurl = {https://ui.adsabs.harvard.edu/abs/2021MNRAS.506.2824L}
}

@ARTICLE{Penucci_etal2014,
       author = {{Pennucci}, Timothy T. and {Demorest}, Paul B. and {Ransom}, Scott M.},
        title = "{Elementary Wideband Timing of Radio Pulsars}",
      journal = {\apj},
         year = 2014,
        month = aug,
       volume = {790},
       number = {2},
          eid = {93},
        pages = {93},
          doi = {10.1088/0004-637X/790/2/93},
archivePrefix = {arXiv},
       eprint = {1402.1672},
 primaryClass = {astro-ph.IM},
       adsurl = {https://ui.adsabs.harvard.edu/abs/2014ApJ...790...93P}
}

@ARTICLE{Lyne_etal1990,
       author = {{Lyne}, A.~G. and {Manchester}, R.~N. and {D'Amico}, N. and {Staveley-Smith}, L. and {Johnston}, S. and {Lim}, J. and {Fruchter}, A.~S. and {Goss}, W.~M. and {Frail}, D.},
        title = "{An eclipsing millisecond pulsar in the globular cluster Terzan 5}",
      journal = {\nat},
         year = 1990,
        month = oct,
       volume = {347},
       number = {6294},
        pages = {650-652},
          doi = {10.1038/347650a0},
       adsurl = {https://ui.adsabs.harvard.edu/abs/1990Natur.347..650L}
}

@ARTICLE{Nice_etal1990,
       author = {{Nice}, D.~J. and {Thorsett}, S.~E. and {Taylor}, J.~H. and {Fruchter}, A.~S.},
        title = "{Observations of the Eclipsing Binary Pulsar in Terzan 5}",
      journal = {\apjl},
         year = 1990,
        month = oct,
       volume = {361},
        pages = {L61},
          doi = {10.1086/185827},
       adsurl = {https://ui.adsabs.harvard.edu/abs/1990ApJ...361L..61N}
}

@ARTICLE{Rosenthal_etal2025,
       author = {{Rosenthal}, Alexandra C. and {Ransom}, Scott M. and {Corcoran}, Kyle A. and {DeCesar}, Megan E. and {Freire}, Paulo C.~C. and {Hessels}, Jason W.~T. and {Keith}, Michael J. and {Lynch}, Ryan S. and {Lyne}, Andrew and {Nice}, David J. and {Stairs}, Ingrid H. and {Stappers}, Ben and {Strader}, Jay and {Thorsett}, Stephen E. and {Urquhart}, Ryan},
        title = "{A 34 yr Timing Solution of the Redback Millisecond Pulsar Terzan 5A}",
      journal = {\apj},
         year = 2025,
        month = apr,
       volume = {982},
       number = {2},
          eid = {170},
        pages = {170},
          doi = {10.3847/1538-4357/adb8cd},
archivePrefix = {arXiv},
       eprint = {2410.21648},
 primaryClass = {astro-ph.HE},
       adsurl = {https://ui.adsabs.harvard.edu/abs/2025ApJ...982..170R}
}

@ARTICLE{Stairs_etal2001,
       author = {{Stairs}, I.~H. and {Manchester}, R.~N. and {Lyne}, A.~G. and {Kaspi}, V.~M. and {Camilo}, F. and {Bell}, J.~F. and {D'Amico}, N. and {Kramer}, M. and {Crawford}, F. and {Morris}, D.~J. and {Possenti}, A. and {McKay}, N.~P.~F. and {Lumsden}, S.~L. and {Tacconi-Garman}, L.~E. and {Cannon}, R.~D. and {Hambly}, N.~C. and {Wood}, P.~R.},
        title = "{PSR J1740-3052: a pulsar with a massive companion}",
      journal = {\mnras},
         year = 2001,
        month = aug,
       volume = {325},
       number = {3},
        pages = {979-988},
          doi = {10.1046/j.1365-8711.2001.04447.x},
archivePrefix = {arXiv},
       eprint = {astro-ph/0012414},
 primaryClass = {astro-ph},
       adsurl = {https://ui.adsabs.harvard.edu/abs/2001MNRAS.325..979S}
}

@ARTICLE{Manchester_etal2001,
       author = {{Manchester}, R.~N. and {Lyne}, A.~G. and {Camilo}, F. and {Bell}, J.~F. and {Kaspi}, V.~M. and {D'Amico}, N. and {McKay}, N.~P.~F. and {Crawford}, F. and {Stairs}, I.~H. and {Possenti}, A. and {Kramer}, M. and {Sheppard}, D.~C.},
        title = "{The Parkes multi-beam pulsar survey - I. Observing and data analysis systems, discovery and timing of 100 pulsars}",
      journal = {\mnras},
         year = 2001,
        month = nov,
       volume = {328},
       number = {1},
        pages = {17-35},
          doi = {10.1046/j.1365-8711.2001.04751.x},
archivePrefix = {arXiv},
       eprint = {astro-ph/0106522},
 primaryClass = {astro-ph},
       adsurl = {https://ui.adsabs.harvard.edu/abs/2001MNRAS.328...17M}
}

@ARTICLE{Madsen_etal2012,
       author = {{Madsen}, E.~C. and {Stairs}, I.~H. and {Kramer}, M. and {Camilo}, F. and {Hobbs}, G.~B. and {Janssen}, G.~H. and {Lyne}, A.~G. and {Manchester}, R.~N. and {Possenti}, A. and {Stappers}, B.~W.},
        title = "{Timing the main-sequence-star binary pulsar J1740-3052}",
      journal = {\mnras},
         year = 2012,
        month = sep,
       volume = {425},
       number = {3},
        pages = {2378-2385},
          doi = {10.1111/j.1365-2966.2012.21691.x},
archivePrefix = {arXiv},
       eprint = {1207.2202},
 primaryClass = {astro-ph.HE},
       adsurl = {https://ui.adsabs.harvard.edu/abs/2012MNRAS.425.2378M}
}

@ARTICLE{Bassa_etal2011,
       author = {{Bassa}, C.~G. and {Brisken}, W.~F. and {Nelemans}, G. and {Stairs}, I.~H. and {Stappers}, B.~W. and {Kramer}, M.},
        title = "{The binary companion of PSR J1740-3052}",
      journal = {\mnras},
         year = 2011,
        month = mar,
       volume = {412},
       number = {1},
        pages = {L63-L67},
          doi = {10.1111/j.1745-3933.2010.01006.x},
archivePrefix = {arXiv},
       eprint = {1012.5254},
 primaryClass = {astro-ph.HE},
       adsurl = {https://ui.adsabs.harvard.edu/abs/2011MNRAS.412L..63B}
}

@article{Stappers_etal2018,
  author = "Stappers, Ben  and  Kramer, Michael",
  title = "{An Update on TRAPUM}",
  doi = "10.22323/1.277.0009",
  journal = "PoS",
  year = 2018,
  volume = "MeerKAT2016",
  pages = "009"
}

@ARTICLE{Dupree1986,
       author = {{Dupree}, A.~K.},
        title = "{Mass loss from cool stars.}",
      journal = {\araa},
         year = 1986,
        month = jan,
       volume = {24},
        pages = {377-420},
          doi = {10.1146/annurev.aa.24.090186.002113},
       adsurl = {https://ui.adsabs.harvard.edu/abs/1986ARA&A..24..377D}
}

@ARTICLE{Ginzburg_etal2021,
       author = {{Ginzburg}, Sivan and {Quataert}, Eliot},
        title = "{Black widow formation by pulsar irradiation and sustained magnetic braking}",
      journal = {\mnras},
         year = 2021,
        month = jan,
       volume = {500},
       number = {2},
        pages = {1592-1603},
          doi = {10.1093/mnras/staa3358},
archivePrefix = {arXiv},
       eprint = {2008.06506},
 primaryClass = {astro-ph.HE},
       adsurl = {https://ui.adsabs.harvard.edu/abs/2021MNRAS.500.1592G}
}

@ARTICLE{Chen_etal2013,
       author = {{Chen}, Hai-Liang and {Chen}, Xuefei and {Tauris}, Thomas M. and {Han}, Zhanwen},
        title = "{Formation of Black Widows and Redbacks{\textemdash}Two Distinct Populations of Eclipsing Binary Millisecond Pulsars}",
      journal = {\apj},
         year = 2013,
        month = sep,
       volume = {775},
       number = {1},
          eid = {27},
        pages = {27},
          doi = {10.1088/0004-637X/775/1/27},
archivePrefix = {arXiv},
       eprint = {1308.4107},
 primaryClass = {astro-ph.SR},
       adsurl = {https://ui.adsabs.harvard.edu/abs/2013ApJ...775...27C}
}

@ARTICLE{Lower_etal2024,
       author = {{Lower}, Marcus E. and {Johnston}, Simon and {Lyutikov}, Maxim and {Melrose}, Donald B. and {Shannon}, Ryan M. and {Weltevrede}, Patrick and {Caleb}, Manisha and {Camilo}, Fernando and {Cameron}, Andrew D. and {Dai}, Shi and {Hobbs}, George and {Li}, Di and {Rajwade}, Kaustubh M. and {Reynolds}, John E. and {Sarkissian}, John M. and {Stappers}, Benjamin W.},
        title = "{Linear to circular conversion in the polarized radio emission of a magnetar}",
      journal = {Nature Astronomy},
         year = 2024,
        month = may,
       volume = {8},
        pages = {606-616},
          doi = {10.1038/s41550-024-02225-8},
archivePrefix = {arXiv},
       eprint = {2311.04195},
 primaryClass = {astro-ph.HE},
       adsurl = {https://ui.adsabs.harvard.edu/abs/2024NatAs...8..606L}
}

%%%%%%%%%%%%%%%%%%%%%%%%%%%%%%%%%%%%%%%%%%%%%%%%%%

% Don't change these lines
\bsp	% typesetting comment
\label{lastpage}
\end{document}